\documentclass[journal,10pt]{IEEEtran}
\usepackage[T1]{fontenc}
\usepackage[utf8]{inputenc}
\usepackage{textcomp,amsmath,amssymb,graphicx,booktabs}
\usepackage{pgfplots}
\usepgfplotslibrary{groupplots}
\pgfplotsset{compat=1.17}
\usetikzlibrary{arrows.meta,patterns,plotmarks}
\usepackage[backend=biber,style=ieee,sorting=none,maxbibnames=6,minbibnames=1,giveninits=true,doi=true,url=false,isbn=false,eprint=false,citestyle=numeric-comp]{biblatex}
\usepackage[hidelinks]{hyperref}
\usepackage{subcaption}
\usepackage{placeins}
\hypersetup{pdftitle={Establishing Standards for Assessment and Reporting of SNSPD Performance},pdfauthor={Ari N. Willner and coauthors}}
\graphicspath{{Figures/}{./}}
\usepackage[
    style=ieee,
    url=false,
    doi=true,
    isbn=false,
    eprint=false
]{biblatex}
\providecommand{\DN}{D_{\mathrm{N}}}
\DeclareUnicodeCharacter{03BC}{\ensuremath{\mu}}
\DeclareUnicodeCharacter{2009}{\,}
\DeclareUnicodeCharacter{2212}{\ensuremath{-}}

\title{Establishing Standards for Assessment and Reporting of SNSPD Performance}

\author{
\IEEEauthorblockN{
Ari N. Willner\IEEEauthorrefmark{1},
Alejandro Simon\IEEEauthorrefmark{1},
Reed A. Foster\IEEEauthorrefmark{1},
Francesca Incalza\IEEEauthorrefmark{1},
Ava A. Brule\IEEEauthorrefmark{1}, \\
Philip Donald Keathley\IEEEauthorrefmark{1},
Karl K. Berggren\IEEEauthorrefmark{1}
}

\IEEEauthorblockA{
\IEEEauthorrefmark{1}Research Laboratory of Electronics, Massachusetts Institute of Technology
}
}

\begin{document}

\maketitle

\begin{abstract}
Superconducting nanowire single-photon detectors (SNSPDs) can be made sensitive to single photons with near-unity detection efficiency, ultra-low dark-count rates, and sub-nanosecond timing jitter over a broad wavelength range. However, many studies report data inconsistently, often emphasizing a single performance metric without considering overall device performance.  This inconsistent data reporting limits the field, as it makes it difficult to determine which device characteristics are most important. Different applications require functionalities that may conflict with one another; for example, longer-wavelength operation may require lower operating temperatures.  Moreover, there is no universally accepted metric for defining the figures of merit of an SNSPD. 
In this work, we amass a dataset representing the state of the art of SNSPD fabrication, including operating temperature, operating wavelength, photon count rates (PCR), and dark count rates (DCR). We fit the PCR curves and suggest using the saturation plateau length to gauge the SNSPD’s sensitivity. We then fit the DCR and compare the PCR and DCR. Using this approach, we establish a quantitative framework to determine the optimal biasing condition for a given device and to determine the device’s sensitivity at that optimal biasing condition, enabling meaningful comparisons across disparate devices. Additionally, we offer concrete benchmarks for measuring device performance.
\end{abstract}

\section{Introduction}
\label{sec:introduction}
\IEEEPARstart{C}{omparing} superconducting nanowire single-photon detectors (SNSPDs) requires more than a list of maximum efficiencies and minimum dark-count rates: these values may occur at different bias currents, temperatures, or wavelengths. SNSPDs have become a leading technology for low-light detection because they can combine high detection efficiency (DE), low dark-count rate (DCR), picosecond-scale timing jitter, fast reset time, and broad spectral sensitivity \cite{hadfieldSinglephoton2009, natarajanSuperconducting2012, venzaResearch2025, reddySuperconducting2020, korzhDemonstration2020, marsiliEfficient2012, taylorLownoise2023}, and these advantages have made SNSPDs central to quantum information applications, time-resolved spectroscopy, and astronomy \cite{hadfieldSinglephoton2009, natarajanSuperconducting2012, esmaeilzadehSuperconducting2021}. However, as the field has matured, device comparisons have become increasingly difficult because different papers use disparate operating conditions, not-well-established definitions of efficiency, and inconsistent reported headline values.

The challenge in comparing single-photon detectors is not new \cite{chockalingamSuperconducting2008}. Hadfield's review of single-photon detectors for optical quantum information applications emphasized that detector performance is intrinsically multidimensional, involving detection efficiency, dark-count rate, timing jitter, dead time, photon-number resolution, and spectral range \cite{hadfieldSinglephoton2009}. Later SNSPD-focused reviews similarly organized progress around system detection efficiency, dark counts, timing jitter, reset time, operating temperature, and wavelength coverage \cite{natarajanSuperconducting2012,esmaeilzadehSuperconducting2021}. Other reviews have proposed or discussed more compact figures of merit for single-photon detectors, including ratios that combine efficiency, dark counts, and timing uncertainty \cite{youSuperconducting2020,bronziSPAD2016}. These metrics are useful, but they have not yet been applied systematically across many papers. Furthermore, metrics can be applied inconsistently due to variations in the choice of operating point, including bias current.

This distinction matters because SNSPD performance is strongly bias-current dependent. Increasing the bias current generally improves the photon count rate (PCR) or DE until the detector's PCR saturates, called the saturation plateau, but it also increases the DCR as the device approaches its switching current. A single reported efficiency or DCR value therefore does not fully specify device performance unless the operating bias current is also known. Similarly, the presence or absence of a saturation plateau changes how robustly a detector can be operated: a wide plateau allows the detector to maintain high photon response while using a lower bias current, reducing dark counts and improving stability.

We look to the metrics used in conventional photodetectors for inspiration for SNSPD metrics. Detectivity can be used to quantify how well a linear photodetector can distinguish a weak optical signal from noise \cite{jonesFactors1949,clarkjonesDetectivity1952}. It is closely related to the noise-equivalent power (NEP), the input optical power that produces a signal-to-noise ratio of unity in a $1$\,Hz output bandwidth. A lower NEP and a higher $D$ indicate improved weak-signal sensitivity. However, detectivity does not fully describe single-photon detection, because an SNSPD produces discrete threshold events rather than a continuous photocurrent \cite{hadfieldSinglephoton2009, natarajanSuperconducting2012, venzaResearch2025}. For this reason, we use a detectivity-like metric as a bias-dependent operating-point criterion rather than as a complete replacement for the full set of SNSPD performance metrics.

In this work, we address the comparison problem by the systematic application of candidate metrics to a literature-derived curve database. We collect and analyze a large set of digitized PCR and DCR curves from the SNSPD literature and fit them using a common model, applying the same metrics across devices, materials, temperatures, and wavelengths.

We quantify the bias-current operating margin using the relative saturation plateau width, $\sigma$, and characterize the bias dependence of dark counts. For devices with both reported PCR and DCR, we introduce a normalized detectivity-like metric $\mathcal{D}^{\star}$ as a function of bias current and identify the operating point that maximizes it. The PCR and DCR at this bias current provide a paired summary of detector performance selected by a common criterion, enabling systematic comparisons that account for the tradeoff between photon response and dark counts. This also enables each detector to be compared at an operating point selected by the same rule, rather than by the convention or emphasis of the original paper.

\section{Methods}
\label{sec:methods}
\noindent Our goal is to develop the detectivity-like metric $\mathcal{D}^{\star}$ over a broad range of SNSPD devices and compare it to $\sigma$. In order to do this, we collect PCR and DCR curves from the literature and fit them to a common model, then perform joint operations on them.

We constructed a literature-derived database of SNSPD PCR and DCR measurements from reports that included bias-current-dependent detector response curves 
\cite{
craiciuHighspeed2023,
iovanNbN2023,
marsiliEfficient2012,
engelTantalum2012,
grotowskiImpact2026,
paulEnhanced2026,
reddySuperconducting2020,
vermaHighefficiency2015,
hampelTungsten2026,
nevzorovMoRe2025,
avitabileSuperconducting2025,
zichiOptimizing2019,
knehrNanowire2019,
azemEnhancing2025,
yangSuperconducting2025,
lambertyInterfacing2025,
khanMonolithic2024,
marsiliHotspot2016a,
taoSinglePhoton2025,
losHighperformance2024,
huFractal2024,
taylorMidinfrared2022,
cirilloSuperconducting2020,
qinSuperconducting2024,
calozOptically2017,
kahlWaveguide2015,
wolffBroadband2021,
jiangSuperconducting2024,
qinThermalproperty2024,
yangSuperconducting2014,
shibataSNSPD2017,
erbeMo2024,
chinaHighly2023,
marsiliDetecting2013,
ravindranActive2020,
colangeloNitrogendoped2026,
chenEfficient2026,
weiBacksideetched2026,
charaevLargearea2020,
korzhDemonstration2020,
colangeloImpedanceMatched2023,
korneevaSuperconducting2014,
jeannicHighefficiency2016,
najafiFabrication2015,
taylorLownoise2023,
xiaoUltralowfillingfactor2023,
vermaSinglephoton2021,
strohauerCurrent2025,
changEfficient2022,
wollmanUV2017,
zhangNbN2022,
baekSuperconducting2011,
gourguesSuperconducting2019,
maDisorder2024,
charaevSinglephoton2024,
vermaHighefficiency2014,
chengEpitaxial2020,
wangSaturated2026,
panMidinfrared2022,
kleinEnhanced2026a,
kuznesofFibercoupled2026,
chiFractal2018,
zhouFabrication2026,
xuMultimodefibercoupled2026, 
incalzaFastRecovery2026}. The database contained 601 digitized curve records spanning 71 source reports. Reports were retained if they had complete information on: (a) operating temperature, (b) operating wavelength, and (c) a saturated PCR curve. After filtering, 597 curves from 57 reports were retained, including 364 PCR curves and 115 DCR curves. 364 PCR/efficiency fits and 105 DCR fits succeeded; 21 PCR fits failed for lack of saturation plateaus, eight DCR traces were rejected, and two DCR fits failed. PCR and DCR curve counts differ because: (1) not all reports contain DCR curves, and (2) one DCR curve can be used to characterize multiple operating wavelengths, while a new PCR curve is needed for each wavelength. For each curve, we recorded the source, material system, wavelength, temperature, and switching current, where reported. Data available only as plots were digitized into two-column CSV files using PlotDigitizer Pro, version 3.3.9\cite{plotdigitizer}.

Only 49 pairs of PCR and DCR curves matched conditions for the same device. This could be due to a number of reasons: (1) some reports plot DCR on a non-normalized bias current axis while they report PCR on a normalized bias current axis, (2) only a paucity of DCR points were reported, or (3) DCR was measured under different operating conditions.

\begin{figure}[t]
  \centering
  \includegraphics[width=\columnwidth]{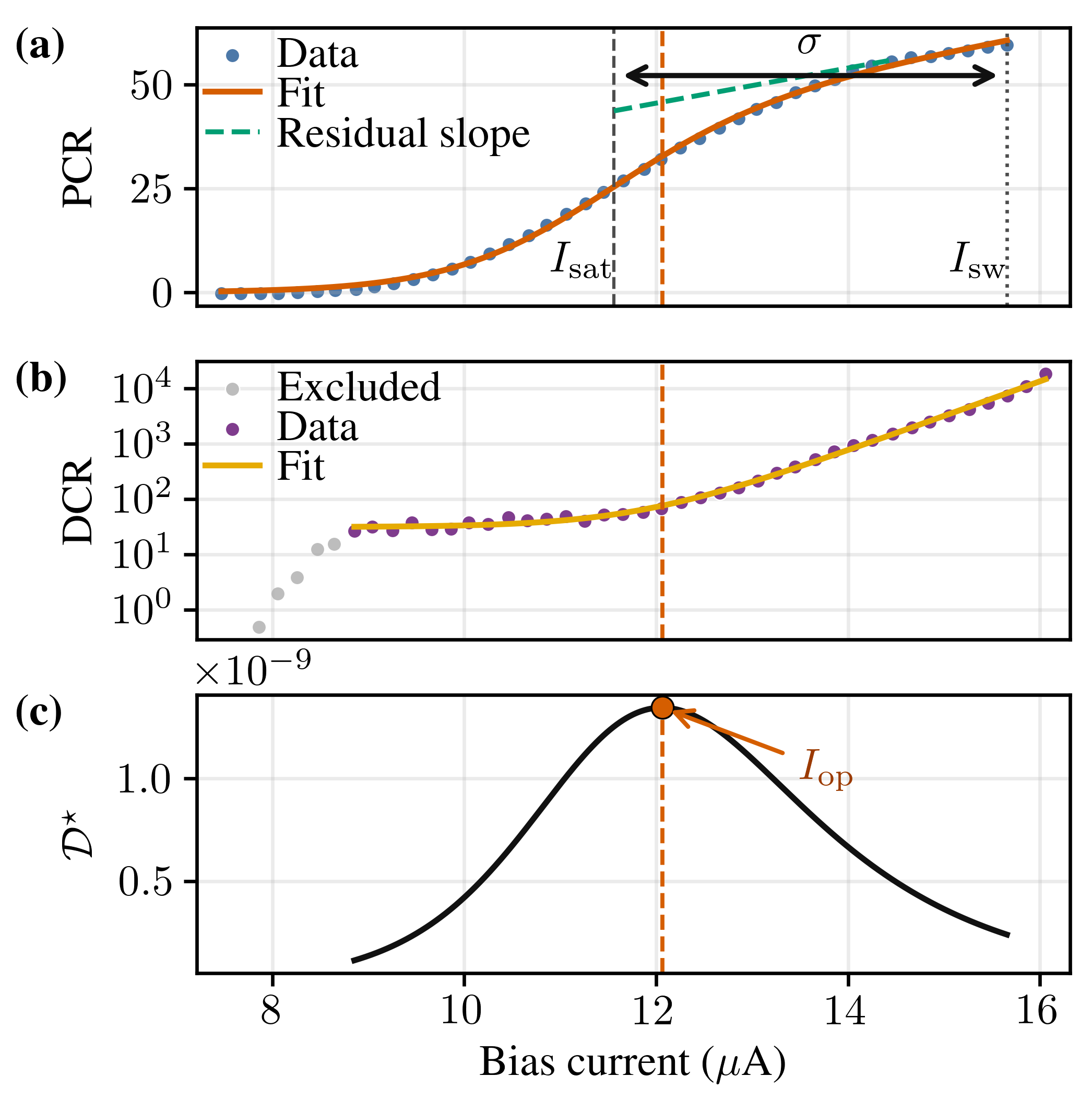}
  \caption{Schematic of the data collection and metric extraction workflow. The data visualized is collected from Xiao et. al. \cite{xiaoUltralowfillingfactor2023}. They measure an SNSPD on NbN at $1550\;nm$ and $2.2\;K$. (a) PCR curve with tanh-line fitting. Saturation plateau width is designated in black. Residual slope is in green, highlighted to justify a line fitting rather than a simple sigmoidal fitting. (b) DCR curve with exponential fitting. Non-physical points at low bias current where the DCR collapses were trimmed. (c) SNR metric calculated from the extracted PCR and DCR curves. The "optimal bias" is when the metric reaches its maximum.}
  \label{fig:methods-combined}
\end{figure}

PCR curves were fit as functions of bias current, $I_b$ (Fig. \ref{fig:methods-combined}). We used a model with a hyperbolic tangent response multiplied by a linear term,
\begin{equation}
\mathrm{PCR}(I_b)
=
\left[a \tanh(b I_b + c) + a \right](d I_b + f).
\end{equation}
The terms $a,\,b,\,c,\,d,$ and $f$ are fit parameters. The linear term $(d I_b + f)$ captures residual slope in the apparent saturation region. The vast majority of the PCR curves obtained an $R^2$ value above $0.99$, indicating a close fitting.

We introduce the relative saturation plateau width, $\sigma$, as a metric for the detector's bias-current operating margin \cite{najafiFabrication2015, strohauerCurrent2025, wolffBroadband2021}:
\begin{equation}
\sigma =
\frac{I_{\mathrm{sw}} - I_{\mathrm{sat}}}{I_{\mathrm{sw}}}.
\end{equation}
The saturation current, $I_{\mathrm{sat}}$, was defined as the bias current at which the fitted PCR curve had the maximum slope. $I_{\mathrm{sw}}$ is the reported switching current or, when unavailable, the largest digitized bias current. Larger $\sigma$ indicates a wider bias margin for maintaining high photon response. 

\begin{figure*}[!t]
    \centering
    \includegraphics[width=\linewidth]{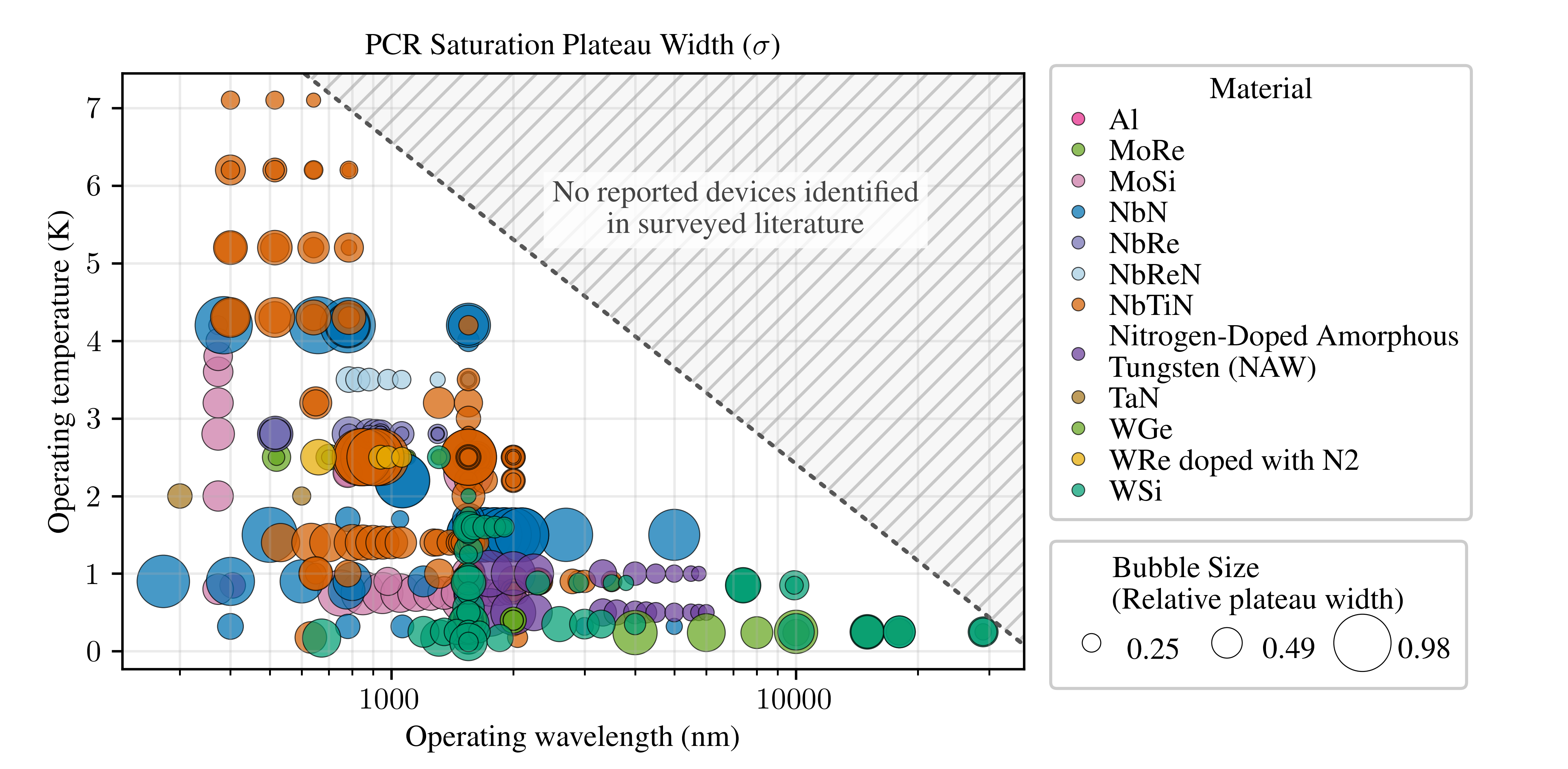}
    \caption{Relative saturation plateau width across operating wavelength and temperature. Each point corresponds to a fitted PCR curve; color denotes material and marker size. Fill fraction denotes relative plateau width. The upper right of the figure shows a region of operation where no reported devices have operated.}
    \label{fig:plateau-width-map}
\end{figure*}

Plateau width alone, however, does not fully determine detector quality: a detector with a broad plateau must still compete with its DCR at the chosen operating point. Therefore, we introduce the normalized detectivity, which combines PCR and DCR data across bias currents.

DCR curves were fit using a floor-plus-exponential model,
\begin{equation}
\mathrm{DCR}(I_b)
=
\mathrm{BCR} + \alpha \exp \left[\beta (I_b - I_{\mathrm{ref}})\right].
\end{equation}
Fitting parameters $\alpha$, $\beta$, and $I_{\mathrm{ref}}$ were used. The floor of the dark count rate is typically associated with the background count rate (BCR), which is not inherent to the device. Rather, it originates from noise in the device setup, such as stray photons. Since the device only operates under a certain range of current bias, the DCR can collapse below the minimum bias needed for detection or above the switching current. Therefore, before fitting, nonphysical low-bias undercounting and high-bias collapse artifacts were removed using an automated trimming procedure. Fits were performed using the logarithm of the count rate to weight the low- and high-count regions more evenly.

PCR and DCR curves were paired when they corresponded to compatible material, temperature, source report, and source device. The default tolerance was 0.05 K in operating temperature. For each accepted PCR--DCR pair, both fitted curves were evaluated over their common current range.

For each pair, we computed a normalized detectivity metric ($\mathcal{D}^{\star}$),
\begin{equation}
\mathcal{D}^{\star}(I_b)
=
\frac{\text{norm}[\mathrm{PCR}(I_b)]}{\sqrt{\mathrm{DCR}(I_b)}},
\end{equation}
where $\mathrm{PCR}(I_b)$ is the fitted PCR and $\mathrm{DCR}(I_b)$ is the fitted DCR. To accommodate raw PCR, normalized PCR, and DE reporting, each fitted PCR curve was normalized so that its sigmoidal component saturates at unity. The maximum value of $\mathcal{D}^{\star}(I_b)$ and the corresponding bias current were recorded.

In order to ensure that the collected data would not be corrupted by a single stray data point, we assessed robustness by refitting each PCR curve 25 times with randomized initial parameters and estimated the error of $I_{\mathrm{sat}}$ and $\sigma$.

\section{Analysis}
\label{sec:analysis}
\noindent The fitted PCR database provides a broad view of saturation behavior across reported SNSPD material systems, wavelengths, and operating temperatures. Of the retained dataset of 129 PCR curves, 122 PCR curves had the wavelength, temperature, material, and fit-quality metadata required for plateau-width comparison.

The plateau-width map (Fig. \ref{fig:plateau-width-map}) shows that broad $\sigma$ are not confined to a single material system or wavelength band. NbTiN and NbN devices dominate the available dataset, but WSi and MoSi devices also occupy regions of large relative plateau width. The pooled map is useful for identifying the available state of the art, but it should not be interpreted as a simple global scaling law because material, geometry, optical stack, and readout conditions vary between reports.

\begin{figure}[t]
    \centering
    \includegraphics[width=\linewidth]{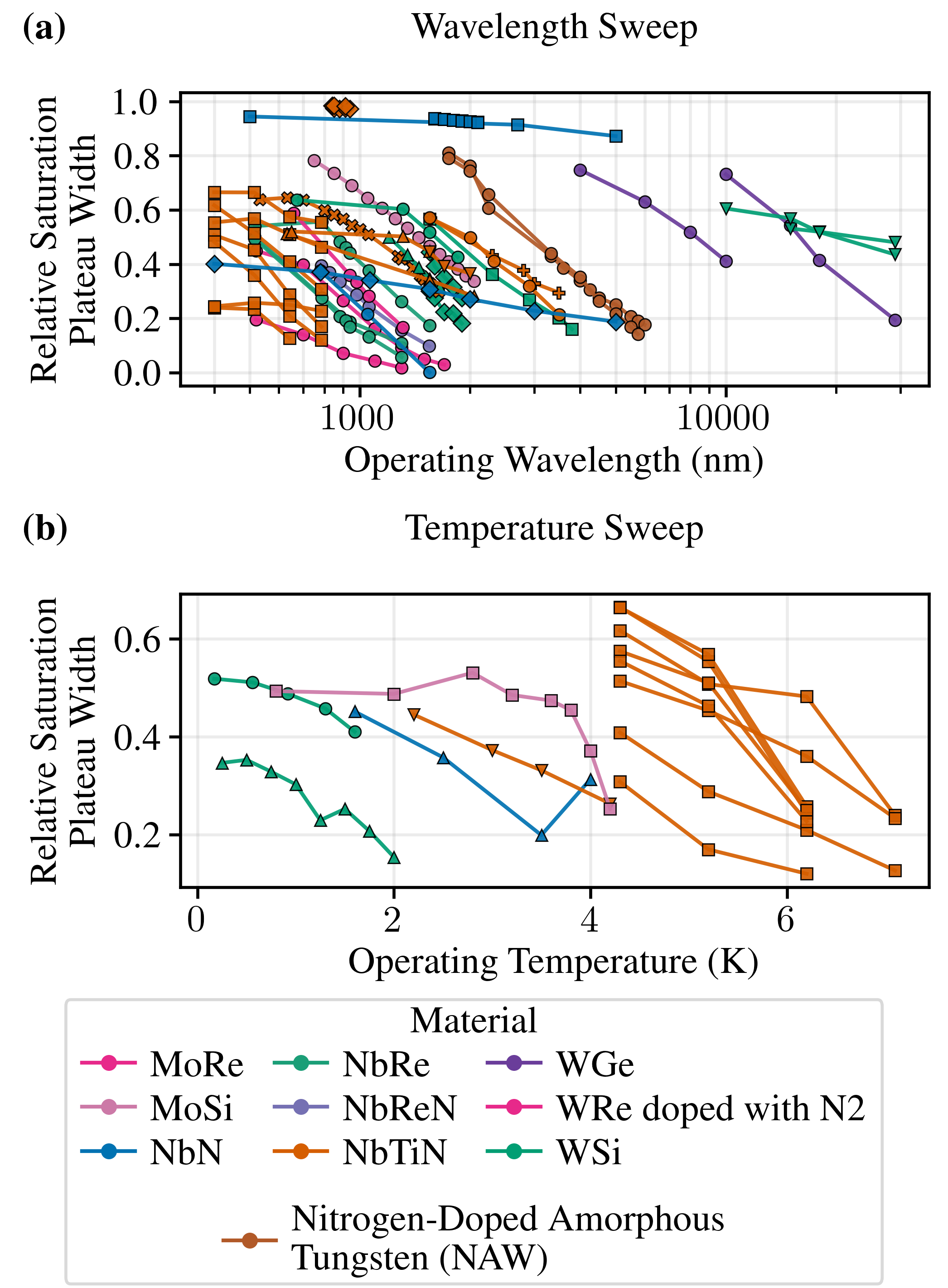}
    \caption{Within-device trends in relative plateau width as a function of operating wavelength and temperature. These plots reduce cross-device variability by comparing sweeps reported for the same or closely related detectors. Color indicates material used; datapoint shape indicates different reference in the literature.}
    \label{fig:device-sweeps}
\end{figure}

To separate device-level trends from cross-paper variability, we also examined curves where the same device was measured over multiple wavelengths or temperatures (Fig. \ref{fig:device-sweeps}). These sweeps show that $\sigma$ generally decreases as operating conditions becomes more demanding, either at longer wavelength or at higher temperature. The dependence of $\sigma$ on temperature and wavelength is highly variable, but there are two general trends: (1) the slope of the decline in the wavelength sweep is fairly consistent between different devices and materials, and (2) there can be a steep fall-off of $\sigma$ at a certain value.

\begin{figure}[t]
 \centering
 \includegraphics[width=\linewidth]{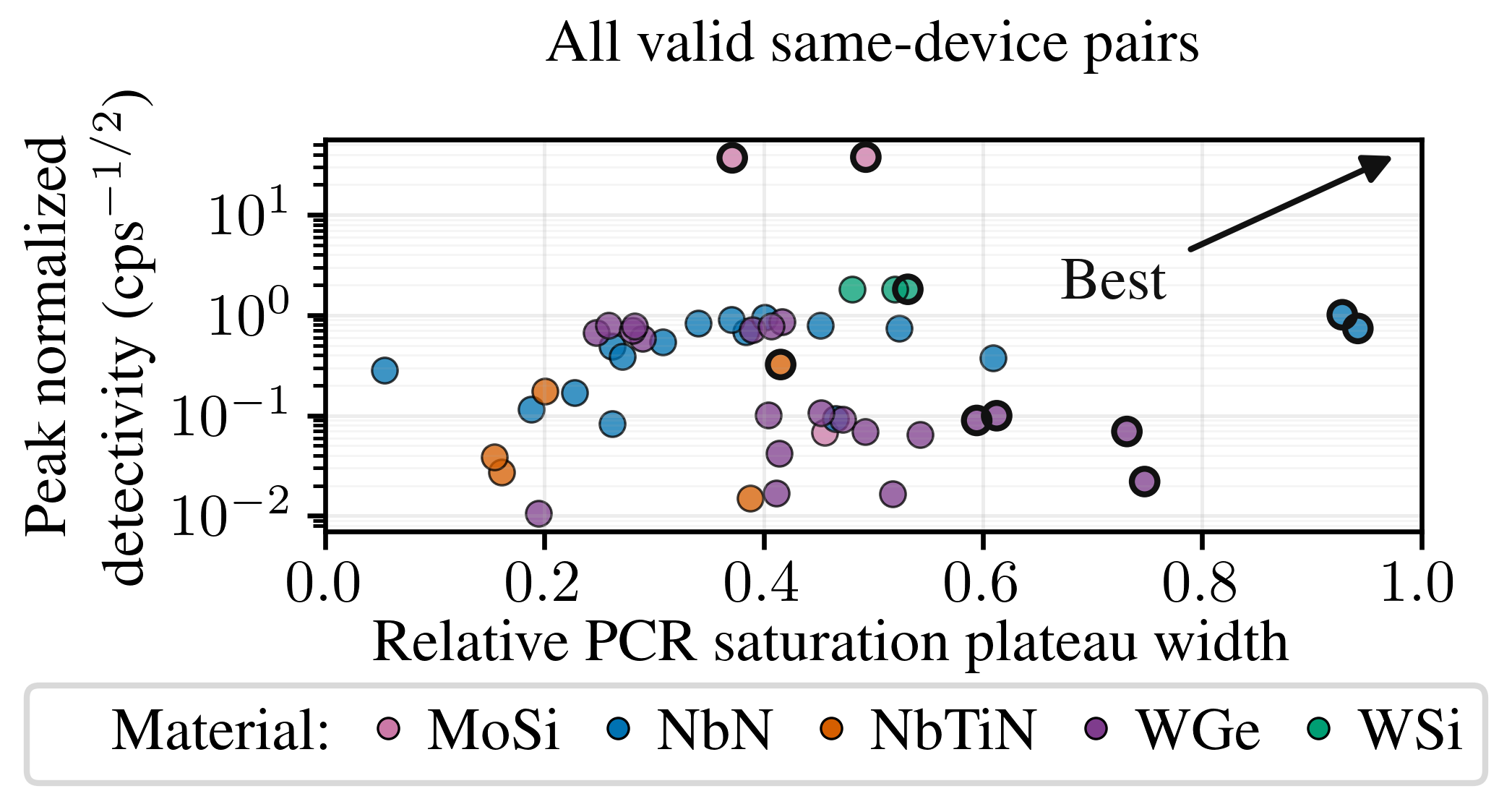}
 \caption{$\sigma$ versus $\mathcal{D}^{\star}$ for all 49 matched PCR-DCR pairs at any wavelength or temperature. Each point uses a same-device PCR-DCR pair with equal recorded temperatures, evaluated over its measured current overlap. Color denotes material. Heavy black outlines identify the ten conditions that are Pareto-optimal across $\sigma$, $\mathcal{D}^{\star}$, higher $\mathrm{PCR}(I_{op})$, and lower $\mathrm{DCR}(I_{op})$ within this pooled set, rather than a frontier defined by the two plotted axes alone. The outlines do not imply a universal ranking across wavelengths and temperatures. The small Spearman correlation ($\rho=0.074$) shows that broad $\sigma$ does not reliably predict high peak $\mathcal{D}^{\star}$ in this sample.}
 \label{fig:plateau-detectivity}
\end{figure}

Figure \ref{fig:plateau-detectivity} compares $\sigma$ with $\mathcal{D}^{\star}$ for all 49 valid matched conditions. The pooled rank correlation is $\rho=0.074$, where $\rho$ is Spearman’s rank correlation coefficient, measuring the strength and direction of monotonic association\cite{CorrelationCoefficientsAppropriate}. Thus, no clear monotonic association is observed in this dataset: a wider fitted PCR plateau does not reliably predict higher peak normalized detectivity. However, this result does not establish statistical independence or exclude associations within particular material families or operating regimes. The sample combines heterogeneous conditions, and its 49 pairs include only 28 device identifiers and 41 distinct DCR traces.

The distinction follows from the quantities entering each metric. Plateau width depends on the PCR transition and the switching-current estimate, whereas $\mathcal{D}^{\star}$ additionally depends on the dark-count curve. Once the photon response changes slowly, a rising DCR can reduce detectivity even as the bias remains within a broad fitted width interval. A broad plateau can provide useful bias margin, but it cannot substitute for a noise measurement.

The highest plotted normalized scores belong to the ultraviolet MoSi measurements of Wollman \emph{et al.}\cite{wollmanUV2017}; the very broad NbN width proxies near $\sigma\simeq0.93$--0.94 come from Wei \emph{et al.}\cite{weiBacksideetched2026}.

\begin{figure}[t]
 \centering
 \includegraphics[width=\columnwidth]{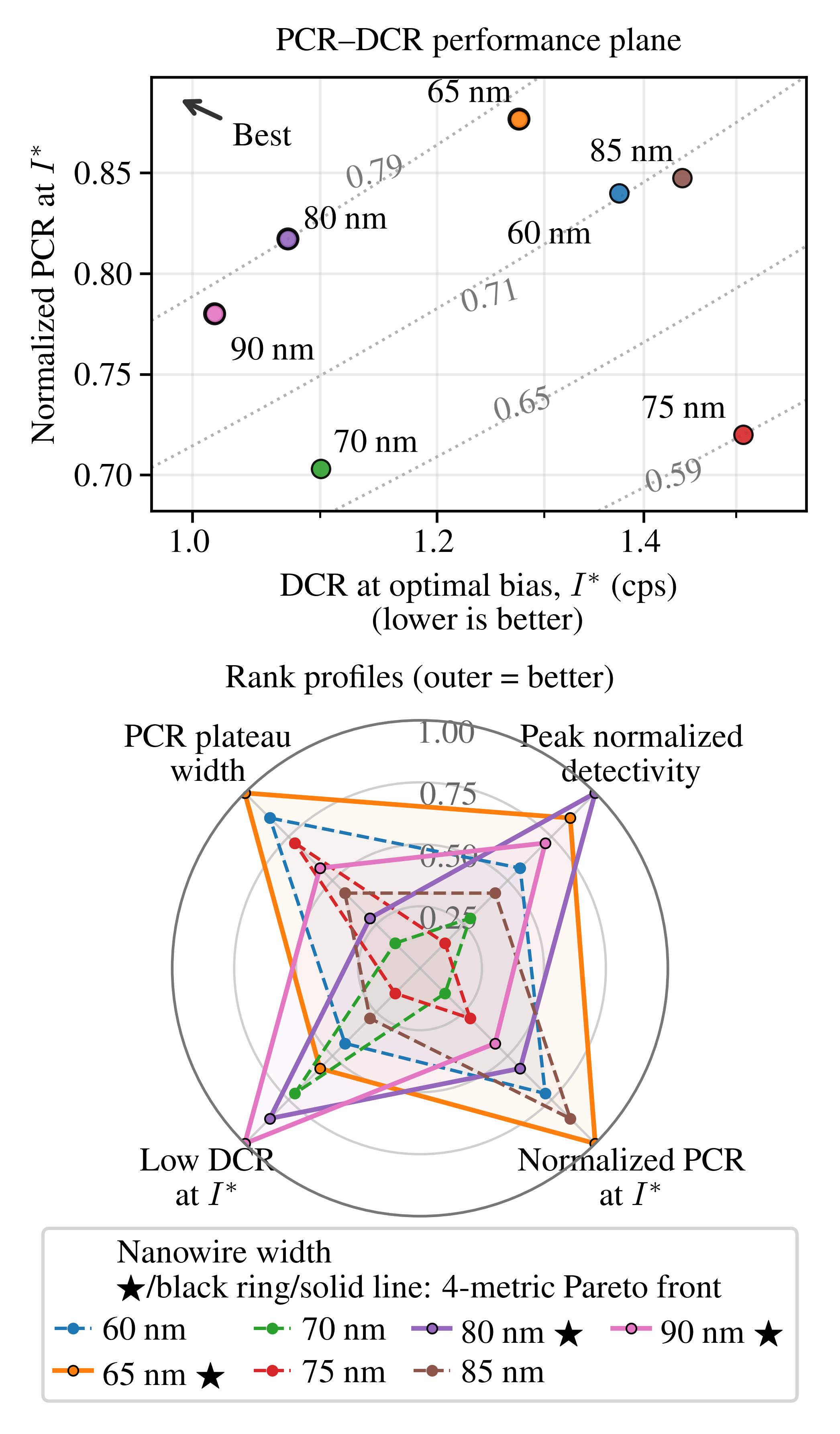}
 \caption{Comparison of seven WGe nanowires with widths of 60-90\,nm from Yang \emph{et al.}\cite{yangSuperconducting2025}, at 2000\,nm and 0.40\,K. Top: $\mathrm{PCR}(I_{op})$ and $\mathrm{DCR}(I_{op})$ at each device's selected bias; the DCR axis is logarithmic and dotted contours indicate equal $\mathcal{D}^{\star}$. Higher response and lower DCR are favored. Bottom: ranks for plateau width, peak normalized detectivity, PCR at $I_{op}$, and low DCR at $I_{op}$. Black rings, solid profiles, and starred legend entries identify the 65, 80, and 90\,nm devices as Pareto-optimal within this seven-device subset using all four displayed metrics. Rank differences and polygon area do not quantify physical performance differences. The 90\,nm optimum occurs at the lower measured-overlap endpoint.}
 \label{fig:spider-comparison}
\end{figure}

The WGe devices of Yang \emph{et al.}\cite{yangSuperconducting2025} provide a more controlled comparison at a common wavelength of $2000$\,nm and temperature of $0.40$\,K (Fig.\ref{fig:spider-comparison}). Seven nanowire widths from 60 to 90\,nm are evaluated at their own $\DN$-maximizing bias currents. The left panel shows the simultaneously attained normalized response and DCR; the right panel shows their ranks together with plateau width and peak normalized detectivity.

The 65\,nm nanowire has the largest width proxy, $\sigma=0.407$, and the highest normalized response at its selected bias, $\mathrm{PCR}(I_{op})=0.877$. The 80\,nm nanowire has a smaller width, $\sigma=0.258$, but the largest normalized detectivity, $\mathcal{D}^{\star}=0.789$\,cps$^{-1/2}$, compared with $0.776$\,cps$^{-1/2}$ for the 65\,nm nanowire. The 90\,nm device has the lowest selected-bias DCR, 1.017\,cps, with $\mathrm{PCR}(I_{op})=0.780$. The seven-device plateau-detectivity rank correlation is $\rho=0.143$, consistent with the lack of a simple ordering in this particular comparison.

The 65, 80, and 90\,nm nanowires are all Pareto-optimal within this subset because their advantages differ. Selection therefore depends on whether the application values response margin, peak score, or false-count suppression most strongly. The 90\,nm optimum is at the lower overlap boundary ($I_{op}=6.198\,\mu\mathrm{A}$), so measurements at lower shared currents would be needed to locate an unconstrained maximum. More generally, exact ranks and Pareto membership should be read alongside numerical differences, uncertainty, and the domain over which each optimum was measured.

\section{Discussion and Conclusions}
\label{sec:discussion}
\noindent A central lesson from this study is that SNSPD performance cannot be fully understood from a single reported efficiency, dark-count rate, or timing value. The relevant operating point depends on the full bias-current dependence of both the photon response and the dark-count response. We conclude that the bias current of maximum $\mathcal{D}^{\star}$ provides a practical criterion for choosing this ideal bias current. The PCR and DCR at that bias current then form a compact summary of device performance. This approach preserves the central tradeoff in SNSPD operation: high photon response is valuable only insofar as it is achieved without an unacceptable increase in dark counts.

Furthermore, one cannot only rely on one of $\mathcal{D}^{\star}$ or $\sigma$ to fully describe their device, and one cannot use $\sigma$ as a proxy for $\mathcal{D}^{\star}$: $\sigma$ indicates the device's robustness to fluctuations in bias current, while $\mathcal{D}^{\star}$ indicates the signal-to-noise ratio. This tradeoff is especially application-dependent. In quantum key distribution, DCR contributes directly to the quantum bit error rate while PCR or detection efficiency sets the sifted-key rate, making low-noise high-efficiency operation, more closely related to $\mathcal{D}^{\star}$, essential at large channel loss \cite{shibataQuantum2014,boaronSecure2018}. In space and deep-space optical links, detector choice is additionally constrained by weak received photon flux, background light, radiation-induced dark counts, timing requirements, and maximum count-rate demands \cite{shawSuperconducting2017, wilsonRadiationInduced2021}. However, other applications may not require such low-dark-count operation, like quantum detector tomography, where signal counts substantially exceed dark counts and a broad $\sigma$ accounting for bias fluctuations might be more valuable than further reductions in DCR \cite{natarajan_quantum_2013}.

In compiling this database, we found that many papers do not report all of the basic metadata and curves needed for this type of comparison. In particular, operating wavelength, operating temperature, PCR versus bias current, and DCR versus bias current are not always reported together for the same device and measurement condition.

We therefore recommend that future SNSPD publications include, at minimum, the following data for each device: the operating wavelength, operating temperature, switching current, PCR or DE curve as a function of bias current, and the full DCR curve as a function of bias current. When a device is claimed to operate over a range of wavelengths or temperatures, PCR curves should be reported across those wavelengths and temperatures rather than at a single representative point. DCR data should also be reported over the same bias-current range whenever possible, since a single dark-count value does not show how quickly the noise rises near the chosen operating bias.

These data should ideally be provided as numerical tables in the supplement or a public repository, not only as plotted figures. Digitizing curves from published plots introduces avoidable uncertainty and makes it difficult to compare devices across papers. Reporting both absolute bias current and normalized bias current, such as $I_b/I_{\mathrm{sw}}$, would also improve comparability between devices with different switching currents.

Finally, we suggest that future papers report a bias-defined operating point in addition to  maximum values. One practical choice is the method used here: select the bias current that maximizes the detectivity-like quantity $\mathrm{PCR}/\sqrt{\mathrm{DCR}}$, then report the PCR and DCR at that same bias current. This gives a more realistic summary of device performance because it accounts for the tradeoff between increasing photon response and increasing dark counts. The maximum PCR alone may occur too close to the switching current to be the best operating condition for many applications. Reporting $I_{op}$, $\mathrm{PCR}(I_{op})$, and $\mathrm{DCR}(I_{op})$ would make SNSPD results easier to compare and would help connect device characterization to realistic system operation.

Further work may include a more robust search of the SNSPD literature, possibly AI-driven, that would create a robust database of all published SNSPD literature. Additionally, we can create theoretical models for the trends indicated above, such as the sudden drop of $\sigma$ at longer wavelengths.

Overall, this work shows that literature-wide SNSPD comparisons are possible, but only when the underlying PCR and DCR data are available. A more consistent reporting standard would allow future studies to compare materials, geometries, temperatures, and wavelength ranges more quantitatively, and would reduce the field's reliance on isolated headline metrics.

\section{Acknowledgments}
\noindent Initial concept and exploration funded under DARPA SynQuaNon (Award No. HR0011-24-9-0311). Further development funded under QuIC-TAQS: Integrated Lithium Niobate Quantum Photonics Platform (OMA-2137723). Finalization funded under the National Science Foundation (NSF) Center for Quantum Networks (EEC-1941583). This material is based upon work supported by the U.S. National Science Foundation Graduate Research Fellowship Program under Grant No. (2141064). Any opinions, findings, and conclusions or recommendations expressed in this material are those of the authors and do not necessarily reflect the views of the National Science Foundation. The authors thank Evan Golden and Gonzalo Esparza for their insightful edits.
\FloatBarrier
\printbibliography
\end{document}